# PAS: Prediction-based Adaptive Sleeping for Environment Monitoring in Sensor Networks

Zheng Yang[1], Bin Xu[2], Jingyao Dai[1], Tao Gu[3]
[1]Hong Kong University of Science and Technology, Hong Kong
[2]Tsinghua University, China
[3]Institute for Infocomm Research, Singapore
{yangzh, daijy}@cse.ust.hk, xubin@tsinghua.edu.cn, tgu@i2r.a-star.edu.sg

**Abstract**

*Energy efficiency has proven to be an important factor dominating the working period of WSN surveillance systems. Intensive studies have been done to provide energy efficient power management mechanisms. In this paper, we present PAS, a Prediction-based Adaptive Sleeping mechanism for environment monitoring sensor networks to conserve energy. PAS focuses on the diffusion stimulus (DS) scenario, which is very common and important in the application of environment monitoring. Different with most of previous works, PAS explores the features of DS spreading process to obtain higher energy efficiency. In PAS, sensors determine their sleeping schedules based on the observed emergency of DS spreading. While sensors near the DS boundary stay awake to accurately capture the possible stimulus arrival, the far away sensors turn into sleeping mode to conserve energy. Simulation experiment shows that PAS largely reduces the energy cost without decreasing system performance.*

## 1. Introduction

Wireless Sensor Networks (WSNs) are composed of a large number of sensor nodes that are densely deployed either inside the physical phenomenon or very close to it [1]. The distributed sensor nodes organize themselves into a multi-hop wireless network and typically collaborate to perform a common task, such as environment monitoring [6], object tracking [2] and scientific observing [13] etc.

A wireless sensor node, being a microelectronic device, can only be equipped with limited power sources, e.g., batteries. Since wireless sensors are usually intended to be deployed in unattended or even hostile environments, it is almost impossible to recharge or replace their batteries. The lifetime of a sensor node is much dependent on its power consumption. Hence, energy efficiency is of highly concern to the WSN design.

In this paper, we focus on the problem of stimulus detection, which is a common task in an environment monitoring application. In this problem, the stimulus, i.e., a liquid pollutant spreads from the source over a continuously enlarging area. The objective of the monitoring system is to detect the diffused area of stimulus. The applications under this circumstance are usually time sensitive; and the detection delay is a crucial metric to evaluate the performance, especially for emergent events. In addition, to achieve a longer surveillance lifetime, an efficient power management scheme is required. This paper proposes the Prediction-based Adaptive Sleeping (PAS) mechanism, which estimates the process of stimulus diffusion and develops adaptive sensor sleep/wakeup schedules based on the diffusion prediction.

Many existing solutions use the redundancy provided by a high-density deployment of sensors to prolong the lifetime of a WSN. Different from those approaches, PAS shows special suitability to the monitoring of diffusion stimulus (DS) which is characterized by its continuously spreading outward. PAS is based on the idea that the sensors near the stimulus boundary work actively to capture possible spreading events, and the far away sensors can be switched to sleep mode to conserve energy. In PAS, the diffusion process of the stimulus is estimated by sensors near the DS boundary. Each sensor predicts the arrival time of the stimulus based on estimation and adjusts its state adaptively. If it takes a long time for the stimulus to reach a sensor, the sensor can safely sleep longer to get higher energy efficiency. Compared with existing power management approaches, PAS explores the characteristics of such application and is able to achieve higher energy efficiency with tolerable detection delay.

The core idea of PAS is to predict arrival time of stimulus for each sensor node, similar to the SAS approach proposed in [9], which, to the best of our knowledge, is the only existing work that shares similarities with our method. While both approaches aim to reduce energy consumptions by applying the adaptive sleeping mechanism on sensor nodes, PAS makes more accurate prediction on the spreading velocity of the stimulus and by adjusting the parameter of alert time, PAS can get a better tradeoff between detection delay and energy consumption. Our analysis shows that SAS is actually a degenerated case of PAS when the alert time is remarkably decreased. We have conducted extensive experiments to evaluate the efficacy and efficiency of PAS. Our comparative study shows that PAS achieves a good balance between detection delay and energy efficiency.

The rest of the paper is organized as follows. Section 2 discusses related works on power management mechanisms. Section 3 describes the principle and design of PAS. Section 4 presents our evaluation results. Finally, we conclude this paper and discuss future work in Section 5.

## 2. Related Work

In recent years, many WSN systems have been developed to support environment monitoring applications [3, 6-8, 12]. However, these systems aim to provide real-world implementation experiences, and they do not focus on the efficient usage of WSNs, neither power efficiency nor monitoring efficacy.

Many algorithms have been proposed to exploit efficient sleep/wakeup schemes for prolonging the surveillance lifetime of a WSN. While most of the algorithms [5, 11, 14, 17] maintain a small set of active sensors without losing connectivity, some algorithms [4, 16] provide adequate terrain by expanding sensor coverage for different application scenarios.

Chen et al. [5] propose Span, a power saving topology maintenance algorithm, which adaptively elects coordinators from all nodes to form a routing backbone and turn off other nodes' radio receivers most of the time to conserve power. Nodes can elect coordinators locally and change their operating role adaptively between coordinator and non-coordinator, hence, Span achieves good network connectivity under a balanced usage of sensor nodes in the network. Without a significant loss of network capacity, Span aims to minimize the number of elected coordinators. Hence, it can achieve a long lifetime.

Ye et al. [16] propose PEAS, which extends the lifetime of a WSN by keeping only a necessary set of sensors working when the density of node deployment is much higher than the necessary density. The key objective of PEAS is to preserve the network sensing coverage, and node failures are considered norms rather than exceptions. In PEAS, sleeping nodes wake up once in a while to probe their neighbors and replace any failed node as needed.

SAS [9] proposes similar ideas to our approach for the DS detection application. It employs a simple method for the local velocity estimation. Our analysis shows that SAS is actually a degenerated case of PAS.

PAS differs with the above approaches in that we aim to explore the spreading process of DS and develop an effective and energy-efficient sleep/wakeup scheme for environment monitoring applications. It provides complements to existing systems.

## 3. Prediction-based Adaptive Sleeping

This section presents the details of PAS. We first provide an overview of the system, and then we explain the sensor operations, the estimation algorithm and the PAS algorithm.

### 3.1 System Overview

For DS monitoring, it would be the ideal case if we can control sensors sleeping time so they wake up at the right moment, just before the arrival of stimulus. In this case, the sensors can capture the arrival of the stimulus accurately while minimizing energy consumption. However, in practice, it is impossible to get the ideal case. A simple intuition is that we can keep the sensors near the stimulus boundary active and make the sensors which are far from stimulus inactive. Based on this idea, in PAS, the sensors along the stimulus boundary cooperate to estimate the spreading velocity, and each sensor predicts the arrival time of stimulus based on the estimations. A sensor can then adjust its state based on the predicted value. A sensor will turn to sleep mode if the predicted arrival time is larger than a threshold to conserve power; otherwise, it should stay active for the upcoming stimulus.

An important task in PAS is to determine how far the stimulus is from a sensor node. We use the arrival time of stimulus as the crucial parameter rather than the space distance. If we can predict the arrival time of stimulus accurately, we will achieve a good performance in terms of delay time and energy consumption.

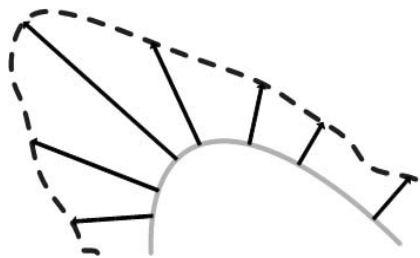

**Figure 1. Stimulus spreading**

A typical case of DS spreading process is illustrated in Fig. 1. The gray line represents the current stimulus boundary and the straight arrows show the spreading velocity at different spots. The DS boundary at the next time will then be the envelope curve of the velocity vectors, represented by the black dashed line. It is obvious that the sensors near the stimulus will be covered by the stimulus shortly while the sensors which are far away from the stimulus will not be affected. In this case, using distance measurement related parameters, e.g., hop count, to determine the arrival of the stimulus may not accurate. In PAS, we use the expected arrival time to estimate if a sensor node is near or far away from the stimulus. Once a sensor calculates the expected arrival time and the value is less than some predefined threshold value, it will remain active. Otherwise, it will go to sleep.

Similar to SAS, each sensor in PAS exchanges the DS information with its neighbors. PAS can achieve more accurate predictions. This is because PAS allows the DS information to be exchanged in a larger field of sensors than SAS, i.e., the sensors which are not covered by the stimulus also transmit alert information, which helps distribute the estimations.

### 3.2 Sensor Operations

There are three types of states for a sensor node in PAS:
- *Covered state*: Sensors get into this state after they detect the stimulus.
- *Alert state*: Sensors switch into this state when the expected arrival time of the stimulus is less than the threshold value. Sensors in this state are usually near the DS boundary.
- *Safe state*: Sensors stay in this state when they have not been notified about the stimulus or the expected arrival time of the stimulus is larger than the threshold value. Sensors in this state are usually far away from the DS boundary.

Sensors in *covered* state and *alert* state should be active to keep on monitoring the area. Sensors in *safe* state may turn to sleep mode to conserve energy. A typical distribution of PAS sensors is shown in Fig. 2. The ALERT area is an irregular shape rather than a circle because the spreading rate of the stimulus may vary in different directions.

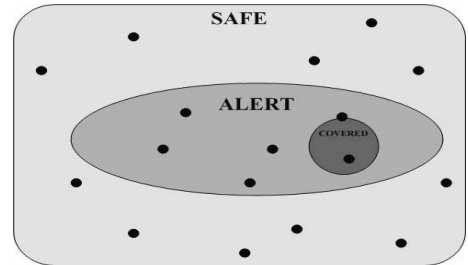

**Figure 2. Sensor statuses**

The state transition diagram of a sensor node in PAS is shown in Fig. 3. All sensors are initially in *safe* state. A sensor will change from *safe* or *alert* state to *covered* state when it detects the stimulus. A sensor in *safe* state will change into *alert* state when its expected arrival time of the stimulus is less than the threshold value. Otherwise, a sensor in *alert* state will go back to *safe* state if its expected arrival time is larger than the threshold value. When the stimulus moves away from a *covered* sensor, the sensor will wait for a detection timeout, and then returns to *safe* state.

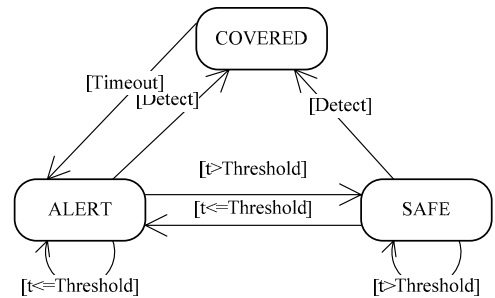

**Figure 3. State transition of a node**

Two types of message are defined to be exchanged among the sensors in a neighborhood in PAS:
- **REQUEST:** A sensor sends this message to request its neighbors for stimulus information. This message does not have any payload.
- **RESPONSE**: A sensor sends this message in response to the REQUEST message. The RESPONSE message contains a sensor's location, state, the estimated spread speed and the predicted arrival time of the stimulus.

Sensors which are in different states have different behaviors as follows.

*Covered state***:** A sensor in this state keeps active. When receiving a REQUEST message, it replies with a RESPONSE message.

*Alert state*: A sensor in this state has complex actions. If it detects the stimulus, it first sends a REQUEST message; then it calculates the expected arrival time according to its neighbors' response, and finally it sends a RESPONSE message to deliver the new changes. If a sensor receives a REQUEST message, it sends back a RESPONSE message. If a sensor receives a RESPONSE message, it re-calculates the expected arrival time and replies with a RESPONSE message if the difference between the expectations has changed significantly.

*Safe state*: A sensor in this state stays in sleep mode. When it wakes up, it changes to *covered* state if it detects the stimulus; otherwise, it sends a REQUEST message. It then calculates the expected arrival time according to its neighbors' responses. If the expected arrival time is less than the threshold value, it will change to *alert* state. Otherwise, it stays in *safe* state.

### 3.3 Arrival Time Estimation

To estimate the DS arrival time accurately is crucial in PAS because it determines a sensor's state and its sleeping scheme. Each sensor obtains DS information from its neighbors, processes the information, and eventually obtains its predicted DS arrival time.

A sensor needs to first compute the spreading velocity in order to calculate the expected arrival time. The velocities vary all the time and may be different at different positions and in different directions. Each sensor has its local estimation. The estimated velocity is a vector which contains both direction and value. The calculation is based on the assumption that the stimulus diffusion is perpendicular to stimulus boundary, e.g., stimulus spreads along the normal direction of the boundary. This is a reasonable assumption for diffusion processes as pointed out in [15].

We define two types of velocity and propose two computation methods to compute them.

*Actual velocity*

When a sensor in *alert* state detects the stimulus, it first sends a REQUEST message, and then calculates the spread velocity of the stimulus based on the responses from its neighbors. This velocity is computed based on actual observation, and hence, it is called actual velocity. A sensor usually has more than one neighboring sensors in *covered* state. Thus, a node usually has more than one velocity reported to it. For sensor $X$, we use formula

$$v_X = \frac{1}{n}\sum_{I=1}^{n}\frac{\overrightarrow{IX}}{t_I}$$

to integrate information from all its neighbors and calculate the actual velocity. In formula above, $I$ is a sensor in *covered* state, $t_I$ is the elapsed time between stimulus detection of sensor $I$ and $X$, and $\overrightarrow{IX}$ is the distance between sensor $X$ and $I$.

*Expected velocity*

Sensors in *alert* or *safe* state need to calculate the expected velocity, which is based on estimation rather than observation. The expected velocity of sensor $X$ is given by

$$v_X = \frac{1}{n}\sum_{I=1}^{n}v_I$$

where sensor $I$ is in *covered* or *alert* state and $v_I$ is the *expected* or *actual velocity* reported from sensor $I$.

The sensor obtains an arrival time from each of its neighbors. The value of expected arrival time is simply the minimum of these arrival times, that is

$$t_X = \min_{I}(\frac{\overrightarrow{IX}\cos \angle I}{v_I})$$

where $\angle I$ is the included angle between $v_I$ and $\overrightarrow{IX}$.

### 3.4 Prediction-based Adaptive Sleeping

To achieve energy conservation, we use different sleeping strategies for sensors in different states. First, sensors in *covered* state are active so that stimulus diffusion can be monitored in real time. Second, to minimize detection delay, sensors in *alert* state also keep active so that they can capture possible spreading events in the future. Third, sensors in *safe* state deploy a specified sleeping strategy such as a linearly increasing sleeping time. By this means we put in-situ sensor nodes into active mode for accurate DS monitoring while outward sensors get to sleep for energy conservation.

When a sensor in *safe* state wakes up and receives stimulus information from its neighbors, it calculates the expected arrival time. If the value is less than the pre-defined threshold, the sensor increases its sleeping interval by adding an increment $\Delta t$ and falls back to sleep. Sensors which have a maximum sleeping interval and their sleeping interval will stay when it reaches the upper bound.

The advantage of PAS is that it can easily adjust the size and shape of an alert area according to stimulus diffusion. For example, the spreading of noxious gas in a city is highly emergent. In this case, the alert area should be enlarged to minimize detecting delays. In a less hazardous case, we can reduce the alert area to cut down energy consumption. By greatly reducing the threshold value of alert time, PAS can degenerate into SAS.

## 4. Experiment Evaluation

We conducted comprehensive simulation to evaluate the efficacy and efficiency of PAS. The simulation is based on the hardware characteristics of Telos [10], the popular used wireless sensor platform. We draw a comparative study with SAS.

### 4.1 Simulation Methodology

We evaluate the performance of different sleeping strategies by a series of experiments. A number of sensors are employed to monitor stimulus diffusion in a specified region. Based on the characteristics of Telos, as shown in Table 1, we evaluate detection delay and energy consumption.

**Table 1**

| Active power(mW) | Sleep power(uW) | Receive power(mW) |
|---|---|---|
| 3 | 15 | 38 |
| Transition power(mW) | Data rate(kbps) | Total active power(mW) |
| 35 | 250 | 41 |

We propose two metrics to evaluate the performance of each sleeping algorithms, average detection delay and average energy consumption [9].

Average detection delay is the average elapsed time between the actual arrival time and the time when a sensor just detects it. This metric is an important measure for environment monitoring. There is no delay for active sensors since they can immediately detect the diffusion while sleeping sensors might miss the first arrival time since they are in sleeping state.

Average energy consumption is the average energy consumed by each sensor. It consists of both controllers' and communication energy consumption. Average energy consumption provides a scale to evaluate the energy efficiency of different sleeping strategies.

### 4.2 Detection Delay

In this experiment, we evaluate the relationship between detection delay and maximum sleeping interval, and compare the results with SAS and non-sleeping (NS) sensors. We set up 30 nodes; and each node has a transmission range of 10m. Maximum sleeping interval is the maximum period when a node remains in sleeping state.

Fig. 4 shows that NS sensors have zero delay since they always keep active. The sleeping periods for both SAS and PAS sensors increase linearly until they reach the maximum values. However, PAS sensors have less latency than SAS sensors.

Fig. 5 shows that, in PAS, the average detection delay decreases from 1.73s to 1.5s when increasing the threshold of alert time from 10 s to 30 s. It demonstrates the adaptability of PAS while both PAS and NS do not such ability. With this characteristic, PAS can be easily applied to non-emergent applications as well.

### 4.3 Energy Consumption

Fig. 6 shows the relationship between energy consumption and maximum sleeping interval with the setup of 30 nodes and a transmission range of 10m. As mentioned above, a longer maximum sleeping interval usually results in a longer sleeping period and less energy consumption. In our experiments, NS sensors consume the most power because they never sleep while both SAS and PAS have lower energy consumption. PAS consumes slightly more energy than SAS because a PAS sensor activates not only its neighbors but also some far-away sensors; however, the difference is trivial.

Fig. 7 shows the energy consumption in PAS varies greatly when increasing the threshold of alert time. This ability suggests that PAS can be also applied to energy restricted situations.

## 5. Conclusions and Future Work

In this paper, we propose PAS to conserve energy for environment monitoring applications in WSNs. By exploring the specific features of DS monitoring scenario, the spreading process is estimated and adaptive sleeping is achieved for sensors. By adjusting the alert time, PAS can achieve a balance between detection latency and energy efficiency. Compared to SAS, PAS obtains a better detection latency at a reasonable energy cost. In our future work, we plan to study the impacts of sensor failure and imperfect communication channel.

## Acknowledgements

This work is supported in part by the National Basic Research Program of China (973 Program) under grant No. 2006CB303000.

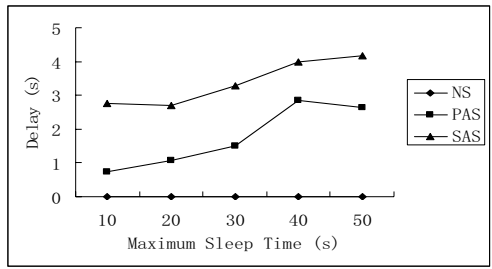

**Figure 4. Detection delay vs. node sleep time**

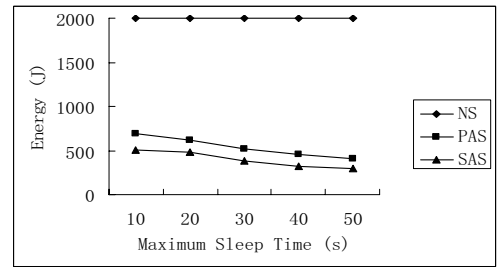

**Figure 6. Energy consumption vs. sleep time**

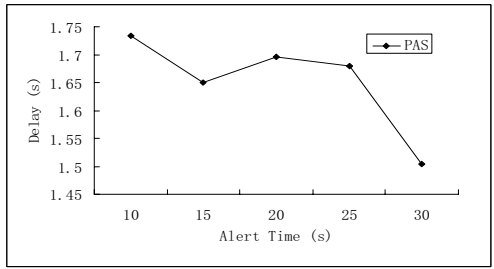

**Figure 5. Detection delay under different alert time threshold sets**

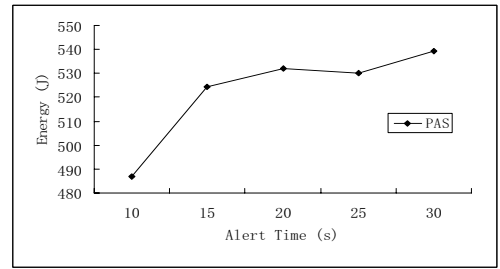

**Figure 7. Energy consumption under different alert time threshold sets**